\documentstyle[epsfig]{europhys}
\input EuroMacr 
\euro{}{}{}{} \Date{} \shorttitle{Diehl et al: Charge renormalization and phase separation in ...}

\date{\today}
\title{Charge renormalization and phase separation in colloidal
suspensions.}
\author{\bf Alexandre Diehl$^{1}$, Marcia C. Barbosa$^{2,3}$ 
and Yan Levin$^{2}$}
\institute{\it  $^1$Departamento 
de F{\'\i}sica, Universidade Federal do Cear{\'a}\\ Caixa 
Postal 6030, CEP 60455-760, Fortaleza, CE, Brazil 
\\$^2$Instituto de F{\'\i}sica, Universidade Federal
do Rio Grande do Sul\\ Caixa Postal 15051, CEP 91501-970, 
Porto Alegre, RS, Brazil\\ 
$^3$Departamento de F\'{\i}sica, Universidade Federal do Rio
Grande do Norte \\ Caixa Postal 66318, CEP  59073-970, Natal, RN, Brazil\\
}
\begin{document}
\pacs{
  \Pacs{82}{70.Dd}{}
    \Pacs{83}{20.Di}{}
      \Pacs{05}{70.Ce}{}
        } 
\maketitle
\begin{abstract}
We explore the effects of counterion condensation on 
fluid-fluid phase separation in
charged colloidal suspensions.  
It is found that formation of double layers around the
colloidal particles stabilizes 
suspensions against phase separation.  Addition of 
salt, however, produces an instability
which, in principle, can lead to a fluid-fluid separation. The instability, 
however, is so weak that it should be impossible to observe
a fully equilibrated coexistence experimentally.
\end{abstract}
\bigskip

Colloidal suspensions present an outstanding challenge
to modern theories of statistical mechanics.  In spite of
the work extending all the way to the beginning of 
previous century, our understanding of these complex
systems remains far from complete.  Even such basic questions 
as what is the form of the interaction potential 
between two colloidal particles inside a solution, still remains 
controversial~\cite{gri00}. The suspensions most often
studied experimentally consist of  polystyrene sulphate spheres
with  diameter in the range of $10^{-6}-10^{-4}$ cm, and
$10^3-10^4$ ionisable surface groups. The typical solvent
is water at room temperaure.  The main stumbling blocks are the presence 
of long range Coulomb interactions and the tremendous asymmetry existing 
between the polyions and their counterions. The ratio of the bare charge 
of a colloidal particle to that of its counterion can be as high as 10000:1. 
This large asymmetry completely invalidates most of the methods of liquid 
state theory, which have proven so successful in the studies of simple 
molecular fluids.  

A particularly interesting question that has provoked much controversy 
over the last two decades concerns itself with the possibility of 
fluid-fluid phase separation in charged colloidal suspensions. A naive 
argument, based on the theory of simple molecular fluids, 
suggests that fluid-fluid (or liquid-gas) coexistence is only possible 
in the presence of sufficiently long ranged attractive interactions. Thus, 
it has been proposed by some authors that a phase separation, or even 
existence of voids in charged colloidal lattices requires attraction
between the polyions~\cite{sog84}. Although appealing intuitively, this 
point of view is difficult to justify within the framework of statistical 
mechanics. The fundamental observation is that colloidal suspension is 
a complex fluid for which many body effects play the fundamental role. 
It is, therefore, erroneous to confine attention  to  pair 
interactions between the colloidal particles while ignoring the significantly 
larger contributions to the free energy arising from the presence of 
counterions~\cite{lev98}. This point has also been 
emphasized by van Roij and Hansen (RH)~\cite{roi97} who demonstrated, in the 
context of the linearized density functional theory, existence of 
``volume'' terms, which can drive phase separation even for 
pairwise {\it repulsive} interactions~\cite{han00}. 
The prediction of a liquid-gas 
phase separation in an aqueous 
solution of like-charged colloidal particles seems, 
however, to be contradicted by the recent simulations of Linse and 
Lobaskin~\cite{lin99}, who did not find any indication of phase 
transition in  suspensions with monovalent counterions. This apparent 
discrepancy between the simulations and the density functional 
theory suggests that a closer look at the mechanism of phase 
separation is worth while. Since the first simulations were 
performed in the absence of salt, as a starting point, we shall 
concentrate our attention on this regime. 

Our model consists of $N_p=\rho_pV$ spherical polyions of radius $a$, 
inside a homogeneous medium of volume $V$ and dielectric constant $D$. Each 
polyion carries $Z$ ionized groups of charge $-q$ uniformly distributed 
over its surface. A total of $ZN_p$ monovalent counterions of charge $+q$ 
are present in order to preserve the overall charge neutrality of solution. 
In the absence of salt, the counterions can be treated as point-like.      

All the thermodynamic properties of colloidal suspensions can be 
determined given the free energy. Unfortunately due to complexity of 
these systems, no exact calculation is possible and approximations must 
be used. We construct the total free energy as a sum of the most 
relevant contributions: electrostatic, entropic, and hard core, 
$f=F/V=f_{el}+f_{ent}+f_{hc}$. The electrostatic free energy, $f_{el}$, 
is the result of polyion-counterion, $f_{pc}$, and the polyion-polyion, 
$f_{pp}$, interactions. Interacions between the monovalent microions 
are insignificant for aqueous solutions 
and can be ignored~\cite{lev98}. 

The polyion-counterion contribution to the total free energy can be 
obtained in the framework of Debye-H\"uckel theory~\cite{lev98,deb23}. 
Fixing one colloidal particle at the origin, it is possible to show that 
the electrostatic potential in its vicinity satisfies the Helmholtz 
equation $\nabla^2 \psi=\kappa^2 \psi$, where 
$\kappa a=(4\pi Z \rho_p^*/T^{\ast})^{1/2}$, and the reduced 
temperature and density are $T^{\ast}=k_BTq^2D/a$ and $\rho_p^*=\rho_pa^3$, 
respectively. The electrostatic free energy can be obtained from the solution 
of the Helmholtz equation followed by the Debye charging process, yielding
\begin{equation}
\label{1}
\beta f_{pc}=  \frac{Z^2}{2T^{\ast}(1+\kappa a)}\rho_p\;.
\end{equation}
We note that this expression would be identical to the one obtained 
by RH using the  
density functional theory \cite{roi97}, but for the self energy contribution, 
$\beta f_{self}=Z^2\rho_p/2T^{\ast}$, which we include and RH excluded. 
Since $f_{self}$ is proportional to the density, it is irrelevant for any 
thermodynamic calculations, {\it as long as} the effects of charge renormalization 
are neglected. Thus, contrary to some earlier claims~\cite{roi99}, the 
Debye-H\"uckel theory is fully consistent with the density functional theory.   

The polyion-polyion contribution to the free energy is calculated within 
the variational approach proposed by Mansoori and Canfield~\cite{man69}. The 
electrostatic DLVO potential~\cite{der41} is used to describe the effective 
pair interactions between the colloidal particles inside the suspension. Based 
on the Gibbs-Bogoliubov inequality $F \leq F_0+\langle U \rangle_0$, Mansoori 
and Canfield replace the free energy by the lowest variational bound. The 
subscript $0$ denotes the reference system of hard spheres whose diameter 
plays the role of a variational parameter. The polyion-polyion free energy 
is given by~\cite{war99}
\begin{equation}
\label{2}
\beta f_{pp}(\eta)=\rho_p \frac{\eta(4-3\eta)}{(1-\eta)^2}+\frac{Z\rho_p}{2}[\lambda^2
G(\lambda)-1]\;.
\end{equation}
The first term is the free energy of the reference hard sphere system, 
while the second is the electrostatic contribution evaluated using the 
Percus-Yevick pair correlation function for hard spheres. The volume 
fraction, $\eta$, plays the role of a variational parameter. The functions 
\begin{eqnarray}
\label{3}
G(\lambda)&=&\frac{\lambda L(\lambda)}{12\eta
[L(\lambda)+\bar{S}(\lambda)e^{\lambda}]}\;,\nonumber\\ 
L(\lambda)&=&12\eta[(1+\frac{\eta}{2})\lambda+(1+2\eta)]\;,\nonumber \\
\bar{S}(\lambda)&=&(1-\eta)^2\lambda^3+6\eta(1-\eta)\lambda^2+18\eta^2\lambda
-12\eta(1+2\eta)\;,
\end{eqnarray} 
are given in terms of $\lambda=2\kappa r_0 \eta^{1/3}$, where  
$r_0$ is a measure of the typical distance between the macroions 
and is given by $4\pi \rho_p r_0^3/3=1$. The variational minimum $\bar{\eta}$ 
is found by solving the equation $\partial f_{pp}/\partial \eta=0$, this has 
to be done numerically. The polyion-polyion contribution to the total 
free energy is 
$f_{pp}(\bar{\eta})$. Again we observe that this calculation is very similar 
to the one performed by RH using the density functional theory. We also note 
that the Eq.~(\ref{2}) already includes the background subtractions emphasized 
by Warren~\cite{war97}.  

\begin{figure}[t]
\vspace*{-2.2cm}
\begin{center}
\epsfxsize=0.45\textwidth
\epsfbox{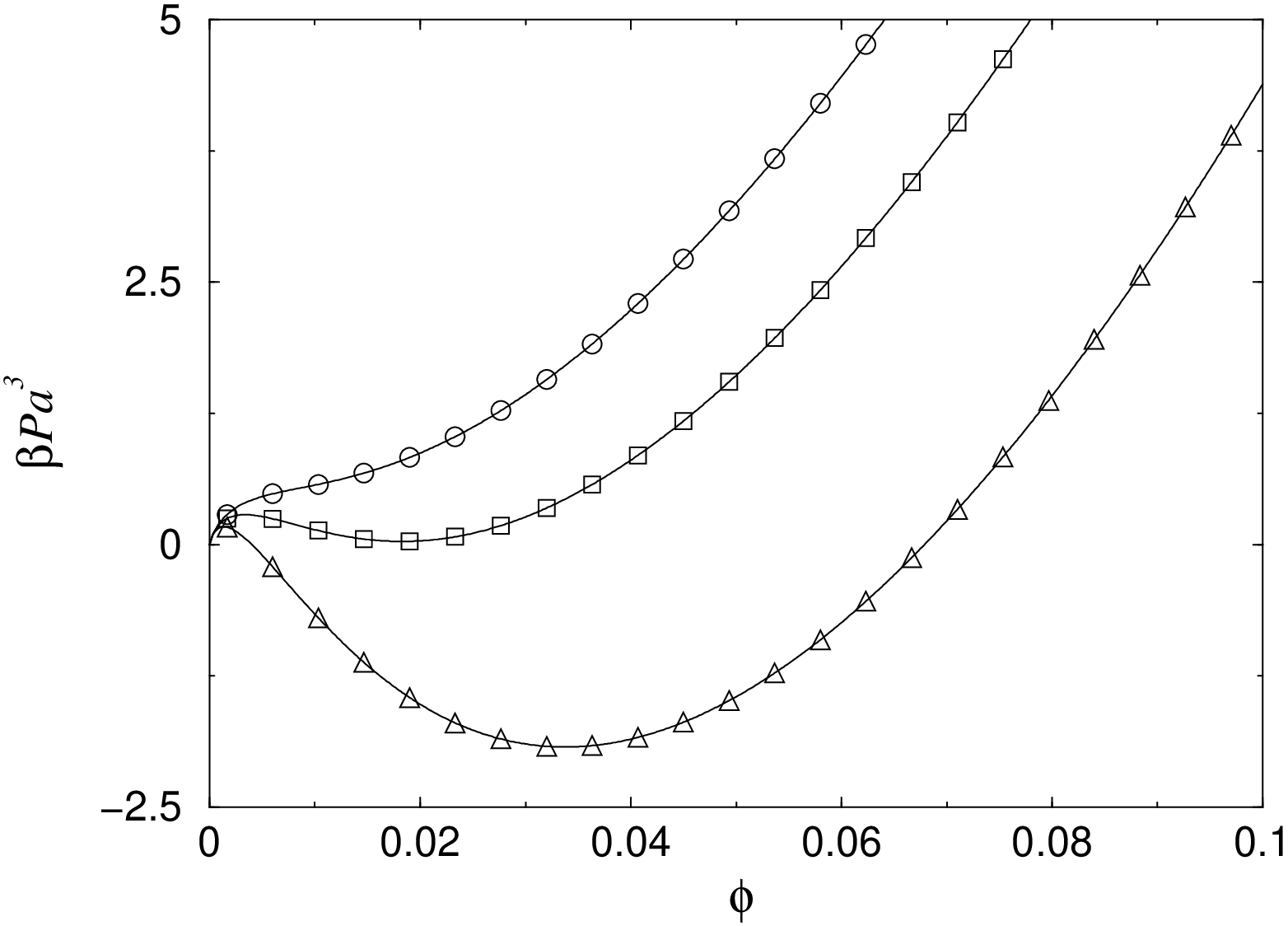}
\end{center}
\vspace*{-.5cm}
\caption{The pressure-volume fraction, $\phi =\frac{4}{3}\pi a^3\rho_p$, isotherms for 
salt-free colloidal suspensions at room temperature, 
$ \lambda_B=7.2\mbox{\AA}$. 
The circles are for $Z=2000$, squares for $Z=2200$ and triangles for $Z=2500$. 
The radius of polyions is $a=1000 \mbox{\AA}$. Note presence of the van der Waals loop 
for $Z=2200$ and $Z=2500$.}
\label{Fig1}
\end{figure}

The contribution to the total free energy arising from the 
hard core repulsion between the
colloidal particles can be approximated by the Carnahan-Starling form~\cite{car69},
\begin{equation}
\label{4}
\beta f_{hc}=\rho_p \frac{\phi_p(4-3\phi_p)}{(1-\phi_p)^2}\;,
\end{equation}
where $\phi_p=4\pi \rho_p^*/3$ is the volume fraction of polyions. 
The final contribution to 
the total free energy is due to entropic motion of counterions and polyions and 
can be expressed using the Flory theory~\cite{flo71},
\begin{equation}
\label{5}
\beta f_{ent}=\rho_p \ln \left(\frac{ \phi_p}{\zeta_p}\right) - \rho_p + Z \rho_p 
\ln \left( \frac{\phi_c}{\zeta_c}\right)-Z \rho_p\;, 
\end{equation}
where $\zeta_p$ and $\zeta_c$ are the internal partition functions of 
the polyions 
and the counterions, and $\phi_c$ is the ``effective'' volume fraction of
counterions. Since both the polyions and the counterions are 
rigid, without any internal structure, $\zeta_p=1$ and $\zeta_c=1$. 
Although we have assumed the counterions to be point-like, the strong electrostatic 
repulsion will prevent them from approaching too close to one another. The 
distance of characteristic approach $d$ can be obtained by comparing the electrostatic 
and the thermal energies. More specifically from the theory of one component 
plasma~\cite{nor84} we find that $d=[(1+3\kappa a/T^{\ast})^{1/3}-1]/\kappa$. 
The volume fraction occupied by the counterions is then $\phi_c=4\pi Z \rho_p d^3/3$. 
Combining all these contributions, we obtain the total free energy 
$f=f_{ent}+f_{pc}+f_{pp}+f_{hc}$ of a colloidal suspension. We observe that 
when $Z\lambda_B/a$ is bigger than approximately $15$, the Helmholtz free energy 
is no longer a convex function of colloidal density 
and the separation into two coexisting phases 
becomes thermodynamically favorable. Here $\lambda_B=a/T^{\ast}$ is the Bjerrum 
length and is equal to $7.2\mbox{\AA}$  in water at room temperature. In Fig.~\ref{Fig1} we 
demonstrate some characteristic pressure-density isotherms, exhibiting the familiar 
van der Waals loop, indicating the presence of a first order phase transition.

Is the phase transition found above consistent with the underlying 
approximations of the Debye-H\"uckel and the 
linearized density functional theory? 
Clearly the fact that the transition is located at $Z/T^{\ast}\approx 15$, 
i.e., the strong coupling regime, should leave us concerned. 
Certainly, in this regime the charge renormalization due to strongly associated 
counterions 
should play a  significant role~\cite{ale84}. Fortunately, it is fairly 
straight forward to include the effects of counterion condensation directly 
into the theory presented above~\cite{lev98,fis93}. 
To achieve this, we separate counterions into 
condensed and free. For simplicity we shall assume that each polyion has an 
equal number $n$ of condensed (associated) counterions~\cite{lev98}. The
density of free  counterions is then $\rho_f=(Z-n)\rho_p$. We shall suppose
that the only effect  of condensed counterions is to renormalize the bare
charge of colloidal particles, while the screening is performed by the {\it
free} microions. The effective  charge of a polyion-counterion complex is then
$Z_{eff}=Z-n$. The free energy $f(n)$,  taking into account the charge
renormalization, is obtained by replacing  $Z \rightarrow Z-n$ in all
formulas, including the inverse  screening length $\kappa$. In addition, since
the complexes now have  structure~\cite{lev98}, their internal partition
function can be approximated  by $\zeta_p=[Z!/((Z-n)!n!)]e^{-\beta E_{n}}$,
where $E_n$ is the electrostatic  energy of $n$ counterions condensed onto the
surface of a polyion. The  electrostatic energy of association can be obtained
through the charging  process~\cite{lev98}, yielding $\beta E_{n}=
-(Zn-n^2/2)/T^{\ast}$.

For fixed volume and number of particles, the equilibrium state of a colloidal 
suspension is determined by the minimum of Helmholtz free energy, 
$f_{ren}=\mbox{min}_n f(n)$. Predictions of this theory for $Z_{eff}$ have 
been recently tested and found to be in a semi-quantitative agreement with 
the experiments~\cite{que00}. We find that the charge renormalization,
Fig. 2b, has a 
most profound effect on the free energy. The $f_{ren}$ is a convex function 
of colloidal 
density {\it for all values of} $Z$, see Fig. 2a. We conclude 
that {\it counterion condensation stabilizes salt-free colloidal suspension 
against phase separation}.

\begin{figure}[h]
\vspace*{0.3cm}
\unitlength=1cm
\begin{picture}(5,5)
\epsfxsize=0.37\textwidth
\put(.5,0){\epsfbox{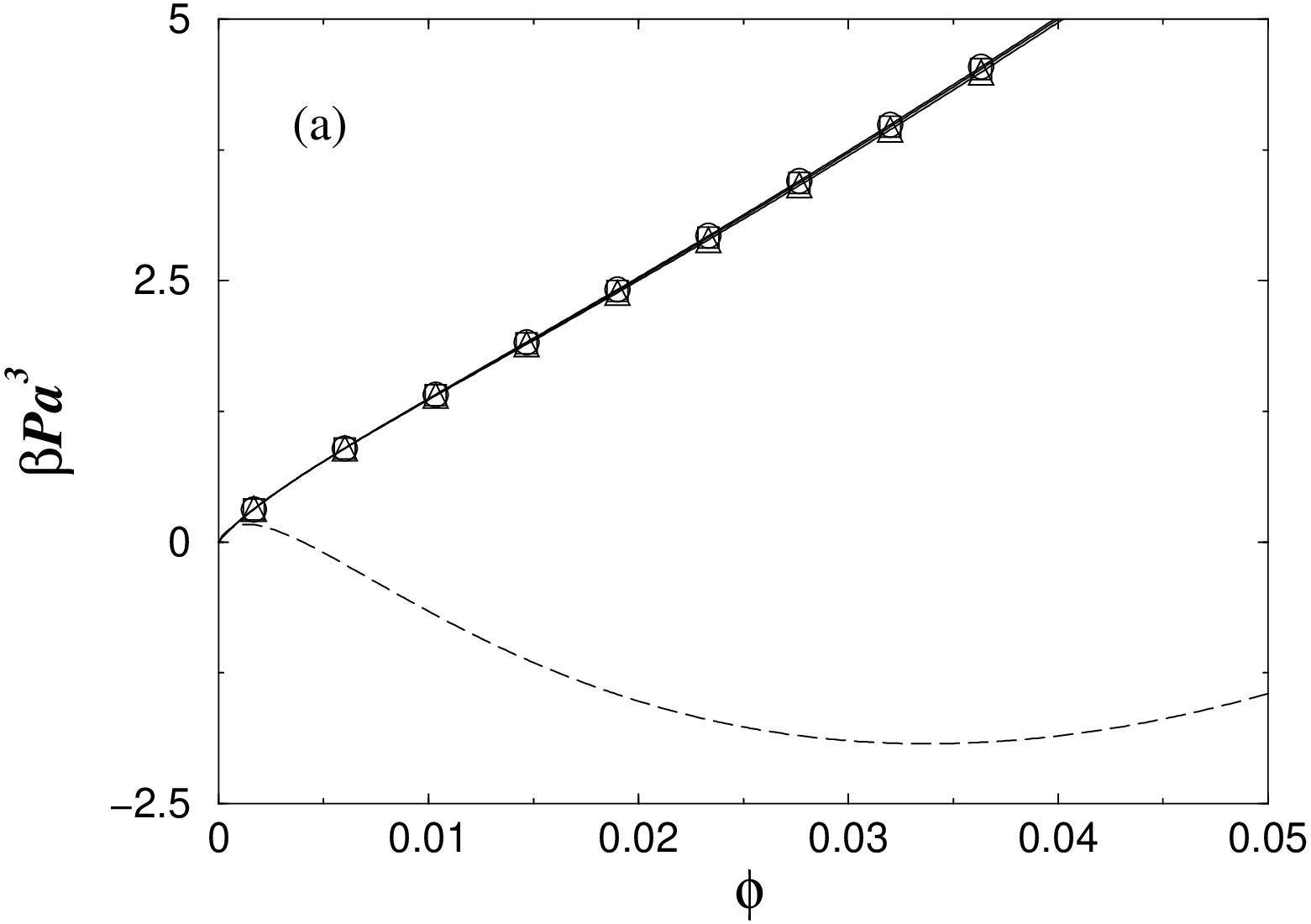}}
\epsfxsize=0.37\textwidth
\put(7.7,0){\epsfbox{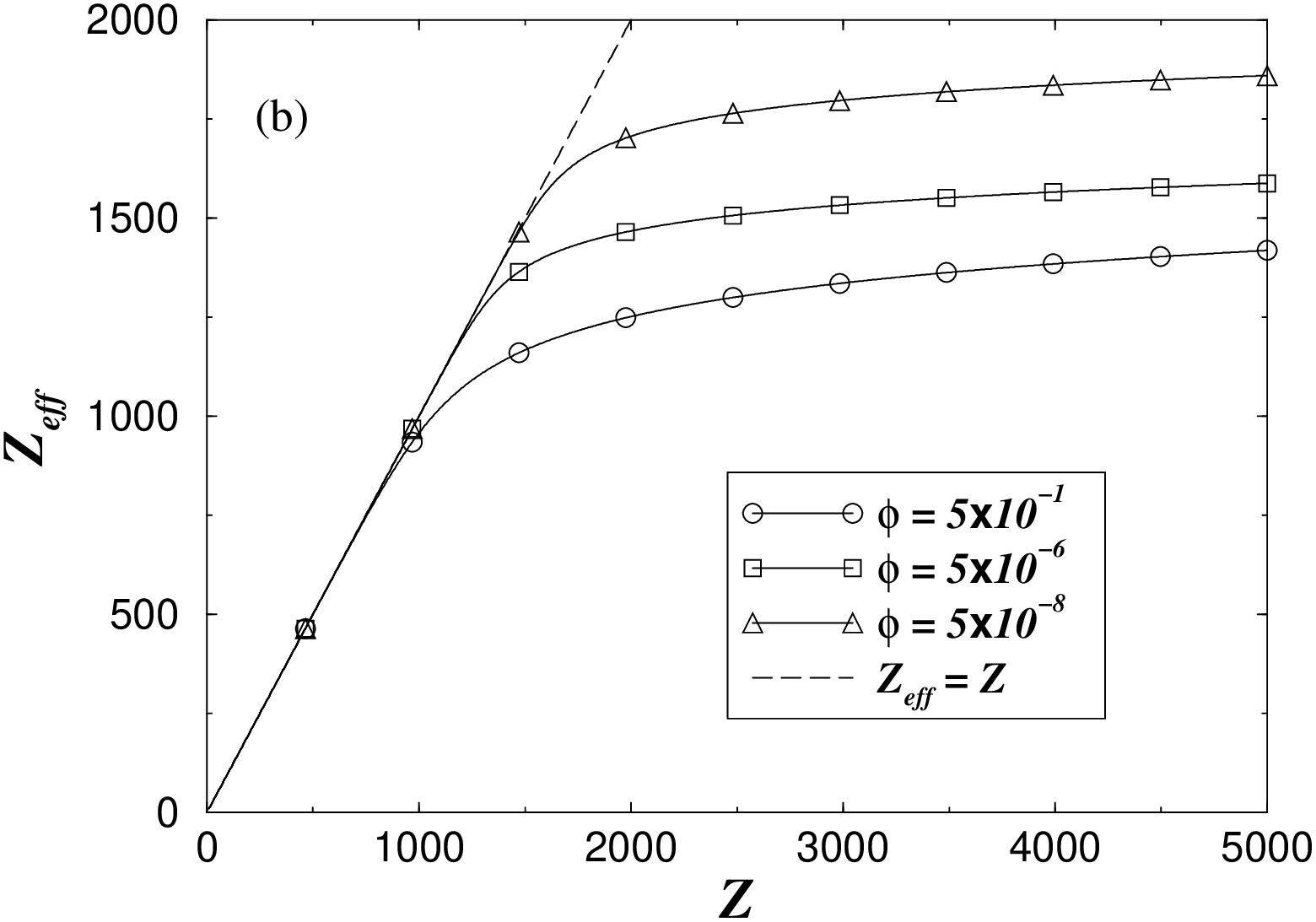}}
\end{picture}
\caption{(a) The pressure-volume 
fraction isotherms for salt-free colloidal 
suspensions. All parameters 
are as in Fig.~\ref{Fig1}. The solid curves are for $Z=2000$ (circles), 
$2200$ (squares), $2500$ (triangles)
{\it with counterion condensation} taken into account.  Note that $Z_{eff}$  
is almost the same for all three bare charges, and the isotherms collapse
onto one monotonically increasing curve. There is no phase separation. 
The dashed curve, included for comparison, {\it neglects} the
counterion condensation,  
and demonstrates the presence of 
the van der Waals loop and the phase separation for $Z=2500$. (b) The
effective charge $Z_{eff}$ vs. $Z$ for various colloidal volume fractions.
Note that the effective charge is quite 
insensitive to  colloidal concentration.}
\label{Fig2}
\end{figure}

We now turn our attention to suspensions in the presence of salt. Since a 
system containing point-like positive and negative particles is intrinsically 
unstable, we assign to each counterion and coion a characteristic radius 
$a_c$. The concentration of monovalent salt is designated by $\rho_s$. The 
calculation now proceeds as the one outlined above. The density of free 
microions is $\rho_f=(Z-n)\rho_p+2\rho_s$ and 
$\kappa a=(4\pi \rho_f^*/T^{\ast})^{1/2}$. The effective charge of a 
polyion-counterion complex is $Z_{eff}=Z-n$. The polyion-counterion and the 
polyion-polyion contributions to the total free energy are the same as for 
the case without salt, but with the new definition of $\kappa$, since all 
free microions contribute to screening. The entropic free energy is given by 
\begin{eqnarray}
\label{6}
\beta f_{ent}=\rho_p \ln \left(\frac{ \phi_p}{\zeta_p}\right) - \rho_p +   
\rho_+\ln \left( \frac{\phi_+}{\zeta_c}\right)-\rho_+ + 
\rho_-\ln \left( \frac{\phi_-}{\zeta_c}\right)-\rho_-\;, 
\end{eqnarray}
where the density of free counterions is $\rho_+=(Z-n)\rho_p+\rho_s$ and the 
density of coions is $\rho_-=\rho_s$. The volume fractions are 
$\phi_+=4\pi \rho_+a_c^3/3$ and $\phi_-=4\pi \rho_-a_c^3/3$.

\begin{figure}[t]
\vspace*{-2.7cm}
\begin{center}
\epsfxsize=0.5\textwidth
\epsfbox{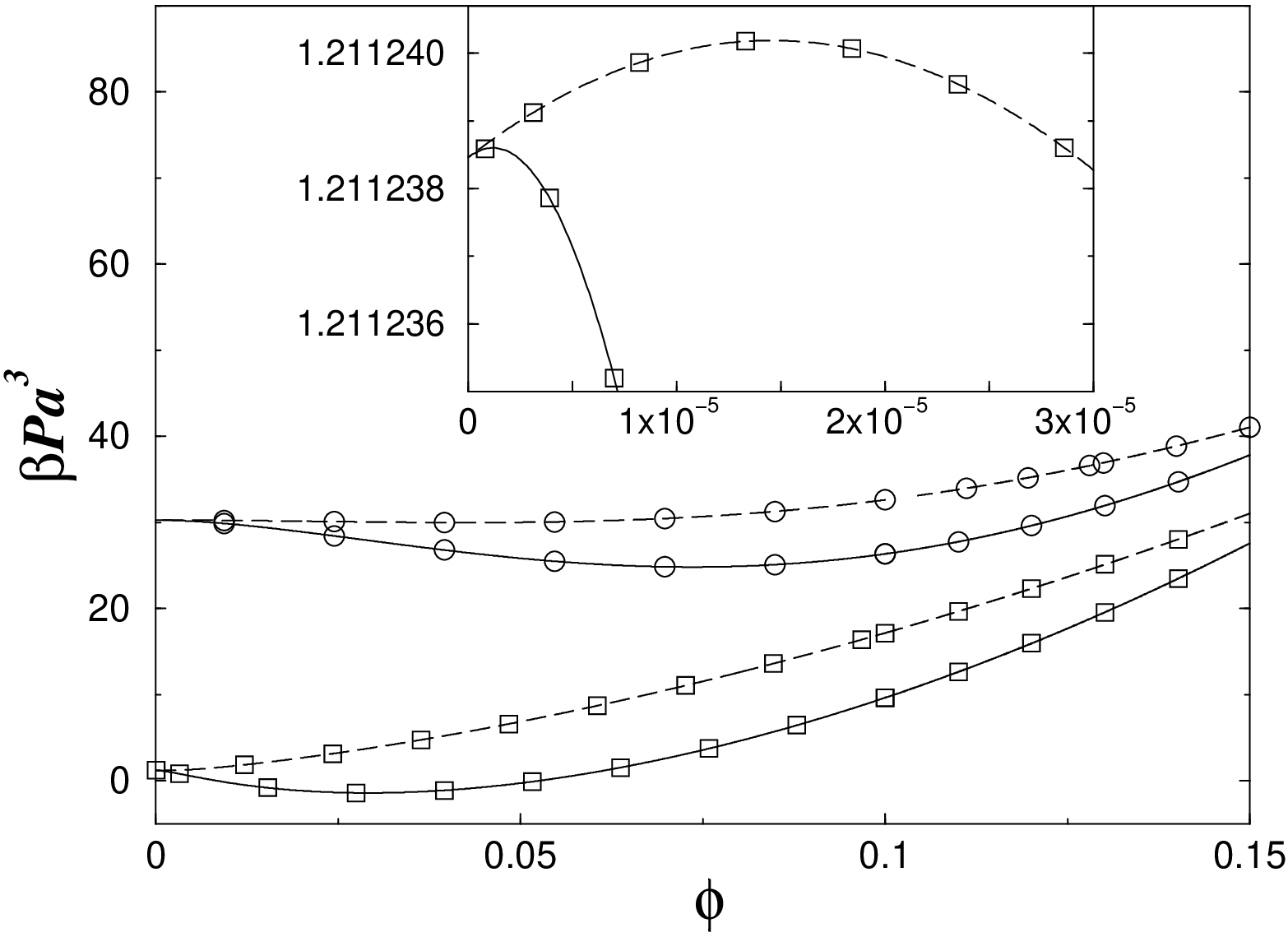}
\end{center}
\vspace*{-.5cm}
\caption{The pressure-volume fraction isotherms for colloidal suspensions 
with two different reservoir salt concentrations, circles  for 
$\rho_s=25 \;\mu M$ and  squares for $\rho_s=1\;\mu M$;
$\lambda_B=7.2$ \AA, $a=1000$ \AA,
$a_c=2$ \AA, $Z=2500$. The solid curves are when the charge 
renormalization is neglected, while the dashed curves are when the charge 
renormalization is taken into account. We note that in the presence of salt, 
the counterion condensation does not fully destroy the phase transition, but only 
makes it weaker. Inset amplifies the region of small densities, showing 
a part of the van der Waals loop for $\rho_s=1\;\mu M$. Note the scale of pressure.}
\label{Fig3}
\end{figure}

To study a possibility of phase separation in this multicomponent system 
is significantly more difficult than for the case of salt-free suspensions. 
The clearest indication of phase transition can still be obtained from the 
pressure-density isotherms. A caution, however, must be taken since the 
two coexisting phases do not necessarily have the same concentration of salt, 
but {\it must
have the same} chemical potential. This can be controlled by
putting the suspension in contact with a hypothetical 
reservoir containing an aqueous solution of salt. 
In Fig.~\ref{Fig3}, we present the 
pressure-density isotherms for  suspensions with salt. 
We see that even an extremely low concentration of salt is sufficient to 
shift the phase transition to the region of parameter space where the charge 
renormalization plays only a marginal role. The suspension phase separates, 
however, the transition is much weaker than it would be in the absence of 
counterion condensation, Fig.~\ref{Fig3}. 
The instability region forms a closed loop
in the $(\rho_s,\rho_p)$ plane.  For the values of
$Z$, $a$, $a_c$, and $\lambda_B$ used in Fig.~\ref{Fig3} 
the upper critical point is at $\rho_s^u\approx 64\mu M$
and $\phi_p^u \approx 0.02$. 
For salt concentrations $\rho_s>\rho_s^u$ 
the suspension is completely stable.  
The lower
critical point is located at $\rho_s^l\approx 0.28\mu M$  and 
$\phi_p^l \approx 0.000025$. Since
in practice it is impossible to deionise water bellow
$\rho_s \approx 1\mu M$, 
all aqueous suspension should --- {\it in principle} --- exhibit 
a miscibility gap at sufficiently low colloidal volume fractions 
and salt concentrations. From Fig.~\ref{Fig3}, however, it is clear 
that fluid-fluid transition is so weak that it is highly unlikely that
a fully equilibrated coexistence can be observed experimentally.  

We conclude that the non-linear effects associated with the presence 
of double layers strongly modify the critical
phenomena of charged colloidal suspensions. 
In particular, we find that  
aqueous suspensions with monovalent counterions 
do not phase separate in the absence of salt.  
Addition
of a small amount of 1:1 electrolyte produces a very weak instability. 
In fact the phase transition 
is so weak that it is  unlikely to  
be observed experimentally.  The metastable
effects associated with the mathematical singularity, however,  
might indeed have 
been detected, appearing as dilute voids in homogeneously deionised
suspensions~\cite{sog84}. 

We would like to stress
that our conclusions are only applicable to {\it aqueous suspensions} with
{\it monovalent} counterions.  In the case of multivalent counterions,
the correlations between the condensed counterions 
can become so strong as to lead to
an effective attraction between the polyions~\cite{ste90}, 
which in
turn can drive a liquid-gas phase
separation. Similarly, in organic solvents of low dielectric 
constant, the correlations between the condensed counterions
can become sufficiently strong as to produce 
phase separation~\cite{lin00}.  
Unfortunately, this correlation induced instability cannot
be studied at the level of the theory presented in this letter
since the interactions between the polyions are
described by the DLVO potential, which does not
include correlations.  
The DLVO potential is sufficient as long as the attention is restricted
to water with monovalent counterions, but most certainly fails
if the electrostatic 
interactions between the condensed counterions are significant, as is
the case for low dielectric solvents and  multivalent counterions.

The theory presented above suggests that the counterion 
condensation stabilizes  charged colloidal suspensions against
fluid-fluid phase separation.  It, however, says nothing
about the possibility that suspension freezes, forming
an ordered lattice.  Fluid-solid transition will
certainly occur at sufficiently large colloidal volume fractions
and  will depend on the concentration of salt in the system. To
study this transition requires an accurate free energy of 
the crystalline state.  This can be obtained
using the density functional theory within a suitable cell geometry
The equality of pressure and  chemical potential in the
two phases will lead to a fluid-solid coexistence curve.

M. C. B. is grateful for the hospitality of the physics department of the 
Universidade Federal do Rio Grande do Norte where part of this work was done. 
This work was supported by the Brazilian science agencies CNPq, FINEP, and Fapergs.

\end{document}